# Astro2020 Science White Paper

# Synoptic Studies of the Sun as a Key to Understanding Stellar Astrospheres

**Thematic Areas**   ☒Stars and Stellar Evolution       ☒Multi-Messenger A&A

**Principal Author:**
Name: Valentin Martinez-Pillet
Institution: National Solar Observatory
Email: vmpillet@nso.edu
Phone: 303-735-7365

**Co-authors:**  F. Hill[1], H. Hammel[2], A de Wijn[3], S Gosain[2], J. Burkepile[3], C. J. Henney[4], J. McAteer[5,6], H. M. Bain[7], W. Manchester[8], H. Lin[9], M. Roth[10], K. Ichimoto[11], Y. Suematsu[12]

[1]National Solar Observatory, Boulder, CO, USA
[2]Association of Universities for Research in Astronomy, Washington DC, USA
[3]High Altitude Observatory/NCAR, Boulder, CO, USA
[4]Air Force Research Laboratory, Space Vehicles Directorate, Kirtland AFB, NM, USA
[5] NMSU, Las Cruces, NM, USA
[6]Sunspot Solar Observatory, Sunspot, NM, USA
[7] CIRES CU Boulder / NOAA SWPC, Boulder, CO, USA
[8] University of Michigan, Ann Arbor, MI, USA
[9] Institute for Astronomy, University of Hawai'i, HI, USA
[10] Leibniz-Institut fur Sonnenphysik (KIS), Freiburg, Germany
[11] Astronomical Observatory, Kyoto University, Japan,
[12] National Astronomical Observatory of Japan, Mitaka, Tokyo, Japan

**Abstract** (optional):

Ground-based solar observations provide key contextual data (i.e., the "big picture") to produce a complete description of the only astrosphere we can study *in situ*: our Sun's heliosphere. The next decade will see the beginning of operations of the Daniel K. Inouye Solar Telescope (DKIST). DKIST will join NASA's Parker Solar Probe and the NASA/ESA Solar Orbital mission, which together will study our Sun's atmosphere with unprecedented detail. This white paper outlines the current paradigm for ground-based solar synoptic observations, and indicates those areas that will benefit from focused attention.

# A multi-pronged approach to understanding an astrosphere

Stars host planets and interact with them using a diverse suite of messengers: best known are gravitational forces and radiation in the form of photons; magnetic fields and charged particles are next in relevance. Three experiments are going to dominate the field of solar and heliospheric physics that use this multi-messenger approach: NSF's Daniel K. Inouye Solar Telescope (DKIST, starting operations in summer 2020), NASA's Parker Solar Probe (PSP) already orbiting the Sun, and the ESA/NASA Solar Orbiter mission.

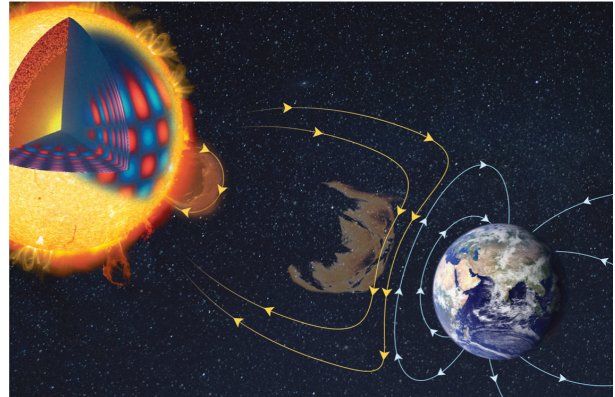

*Figure 1:* Idealization of the magnetic connectivity between the Sun and the Earth. The orientation of the solar magnetic field near the Earth system is referred below as the "Bz" component.

All three projects target the microphysics that describes the magnetic connectivity between the Sun and the heliosphere (**Fig. 1**) in various ways. DKIST will sense with unprecedented detail and sensitivity the physical conditions in the chromosphere and the corona —the layers where the heating and acceleration of the plasma occurs. By flying closer to the Sun, PSP and Solar Orbiter will measure local flows of particles and fields while they still preserve the memory of the processes that originate them back in the solar atmosphere. By using precise magnetic field observations, chemical abundances, and single point *in-situ* field and particle composition measures, these experiments aim at establishing a cause-and-effect relationship that unveils how the magnetic connectivity originates.

These microphysics observations must be put in the context of the large-scale heliospheric conditions. Solar synoptic programs provide this context. However, existing ground based synoptic programs are aging rapidly and are used in ways that differ considerably from their original intentions. This disconnect arises because a wealth of theoretical knowledge with testable predictions has emerged over the last couple of decades (see, e.g., Chen 2017); this knowledge was lacking at the time the synoptic networks' requirements were defined. This white paper describes these new theoretical frameworks behind stellar astrospheres, and discusses the implications for future solar synoptic programs.

Solar synoptic programs also go beyond providing context for the microphysics behind the magnetic connectivity. For example, the Sun's magnetic field is also the driver of transient energy flows between the Sun and all solar system bodies, broadly termed as space-weather. On Earth, space-weather produces the polar auroras and can negatively affect modern technology. On Mars, space-weather processes eroded the atmosphere (Jakosky 2017). With the explosion in exoplanet discoveries has come the realization that space-weather phenomena are also essential inputs to the habitability of planets in general (Garraffo et al. 2017).

Global 3D Magneto-Hydrodynamic (MHD) models of the heliosphere (such as WSA; Wang and Sheeley 1990; Arge et al. 2003) use synoptic magnetic field maps as a basis for modeling a time-dependent description of the background heliospheric solar wind. These simulations are further utilized to emulate the propagation of transient solar wind disturbances such as coronal mass ejections (CMEs) throughout the solar system. Thus, solar synoptic magnetic field maps serve as enabling observation for the simulations they drive, and in turn, our understanding of a vast array of space weather phenomena (i.e. solar wind, co-rotating interaction regions, CMEs and their associated shocks, and the particles accelerated to high energies in those shocks). Much space weather activity can be only be interpreted in light of such global 3D models (Bain 2016).

**Magnetic boundary data that creates the heliosphere**

The Sun-Earth magnetic linkages that define space-weather conditions offer a unique laboratory to observe stellar astrospheres (Wood 2004). Understanding of the resulting coupled systems is a multidisciplinary area of research that has seen significant progress in the last two decades (see, e.g., the reviews by Gombosi et al. 2018 and MacNeice et al. 2018).

The Sun's magnetic field extends to the limits of the heliosphere, where it encounters the interstellar and galactic magnetic fields. Describing the ever-changing magnetic linkages inside the heliosphere starts with the structures and processes at the solar surface. Models such as the WSA describe this magnetic connectivity in near real time. The models include different levels of complexity and represent both the background time-dependent Parker's spiral evolution and episodic phenomena such as CMEs.

All models use similar input data. The background solar wind component starts with synoptic magnetic maps produced regularly by the NSF's GONG synoptic network, operated by the National Solar Observatory (Hill 2018). On the other hand, the explosive CME component often uses a cone-model (Odstrcil & Pizzo 1999) fed by parameters derived from CME observations by space-based coronagraphs (SOHO/LASCO and STEREO). These CME models are purely hydrodynamic. These uses are largely afterthoughts and did not drive the original requirements.

The GONG network was initially designed to measure Doppler velocities to investigate the interior of the Sun. A clever modification in 2002 transformed GONG into a magnetograph network. Its resolution and sensitivity, however, are not optimal for such work: the network has issues with its magnetic zero-point and needs constant monitoring in the reduction pipelines. ***Increased magnetic sensitivity, resolution, and well-calibrated vector magnetic field capabilities are mandatory for improved solar wind modeling.*** This is particularly relevant for the so-called "open flux problem" (Linker et al. 2017) that describes the persistent underestimation of the heliospheric field magnitude predicted by surface magnetograms by as much as a factor of 3.

During the minimum of the solar activity cycle, the coronal holes present at the poles of the Sun establish a quasi-dipolar magnetic configuration in the heliosphere (Owens & Forsyth 2013). Field lines from them extend outwards to all latitudes and form the heliospheric current sheet near the ecliptic. At cycle minimum, the Earth spends most of the time magnetically connected to the polar coronal holes (Luhmann et al. 2009). Unfortunately, these polar coronal holes are the hardest to observe from the Earth (Petrie 2015). In them, the magnetic configuration is relatively simple with vertical field lines, but at the poles this corresponds to *transverse* magnetic field orientations as seen from the Earth. The Zeeman effect makes these transverse signals much harder to observe than the longitudinal ones. Typically, our sensitivity to transverse fields is a full order of magnitude lower. This reduced sensitivity renders the polar coronal holes largely unconstrained in our current modeling efforts. Being the regions where we are magnetically connected for a significant time makes the situation rather unsatisfactory. **To obtain polar maps from the ground that have the required magnetic sensitivity to transverse fields, new approaches with increased Zeeman sensitivity should be pursued.**

## Boundary data and the propagation of solar eruptive structures

In contrast to the open configuration in coronal holes, closed field lines such as active regions are strongly magnetized areas within one solar radius that trap both hot and cool dense plasma. Closed field lines have a smaller impact in the heliosphere unless explosive processes occur in the form of flares or CMEs. Near the center of active regions, magnetic helicity accumulates via unknown processes and leads to the formation of filaments. These structures may eventually eject into the heliosphere in the form of a CME. While filaments have long been known to be strongly associated with CMEs (Munro et al. 1979; Gibson 2018), the accurate measurement of their magnetic topology has become possible only recently (Lites 2014). The photospheric magnetic fields only indirectly map the filament fields. But filaments are key drivers of space weather as they are believed to play a central role in the instability leading to the CMEs (with their material detected at 1 AU, Lepri & Zurbuchen 2010), act as electric current ducts, and store magnetic free energy.

Global heliospheric 3D models propagate CMEs using coronagraph data as input, specifically kinematic and geometric observables of the CME. These models do not propagate magnetized CMEs, and do not predict the value of Bz, the southward magnetic field component in the CME at the Earth – a critical aspect to describe the connectivity between the Earth and the Sun that current models miss (Kilpua, 2017). **Quantifiable models for Bz require the propagation of magnetized CMEs from the solar source into the heliosphere and demands adequate boundary data.** These capabilities represent an area of multidisciplinary research (Jin et al. 2017; Singh 2018; Torok et al. 2018) and are a long-term goal of space-weather investigations.

Mapping the field configuration at the relevant heights will produce quantitative data constraining the filament fields and serve as the initial boundary conditions for the CME propagation. Synoptic observations currently performed at the Dunn Solar Telescope (Sunspot,

NM) by the New Mexico State University consortium include observations of filaments using an He I line and serve as a testbed for new networks.

Compared to the background solar wind conditions, the modeling of the heliospheric propagation of magnetized clouds ejected from the Sun after a CME/flare is in its infancy, and much research is needed (Manchester et al. 2017). During and after the eruption, we expect interactions with the background corona that can produce CME deflections (Kay et al. 2013). In spite of this complexity, some persistent correlations demonstrate that CMEs maintain significant memory of their parent solar region. For example, Marubashi et al. (2015) found that from a sample of 55 CMEs with well-identified solar progenitors and in-situ counterparts, the flux rope orientation measured at 1 AU coincided with the surface polarity inversion line to better than 30°. A similar memory was found in studies by Yurchyshyn et al. (2001) and Palmerio et al. (2017, 2018). ***Such work clearly indicates that data-driven propagation of magnetized CMEs is realistic and only requires the use of proper boundary data.***

## Helioseismology as a window to the sources of solar magnetism

The origin of the solar magnetism described above lies in the interior below the visible photosphere. This is where large-scale motions of the plasma generate the field via, presumably, a dynamo mechanism. Helioseismology—analysis of properties of acoustic waves traveling in the solar interior—is the best probe of these motions.

While helioseismology has succeeded in determining the large-scale flows averaged over the entire Sun, it has not been successful in reliably determining the structure and dynamics below magnetized regions. For example, wave-speed perturbations beneath an active region obtained from time-distance inversions are different from ring-diagram inversions (Gizon et al. 2009) and disagree with the results from semi-empirical models (e.g., Moradi et al. 2010). This clearly indicates inadequate treatment of wave behavior near strong and inclined magnetic fields during the helioseismic analysis. ***The missing ingredient is currently thought to be the transformation of the acoustic waves used for helioseismology into additional MHD oscillation modes (Alfvén waves, fast magnetic, slow magnetic, etc.) at the height in the atmosphere where β, the ratio of thermodynamic gas pressure to magnetic pressure, equals 1.*** This level marks the boundary between atmospheric regions where the magnetic field dominates the plasma motions (β < 1), and regions where the gas pressure is the primary driver (β > 1).

There has been substantial theoretical and observational work on this problem in the last decade (e.g., Cally, Moradi & Rajaguru 2016; Tripathy et al. 2018). These studies have sought a method of correcting the helioseismic analyses in active regions. While there may well be further improvements, the current approach is to forward model the effect using a sunspot background model with embedded acoustic sources (Moradi et al. 2015). This produces estimates of the perturbation to the travel time of waves passing through the region as a function of field strength, inclination, azimuthal angle, height, input wave frequency and travel distance. These perturbed travel times can then be applied to the relevant time-distance helioseismic analyses. With additional theortical work plus routine multi-wavelength multi-height helioseismic observations, we may reach the long-standing promise of detecting active regions on the Sun before they emerge to the surface.

Helioseismic holography is a technique used to infer active regions on the far-side surface of the Sun (Lindsey & Braun 2000, 2017). This approach exploits acoustic travel-time reductions in magnetized areas that result in a phase shift of the waves. This makes large active regions readily apparent in seismic travel-time images reconstructed using observations of global p-modes (**Fig 2.**). Since the sensitivity in these maps depends on accurate and precise measurements of the phase shift between acoustic waves in the solar atmosphere, the improved understanding of phase shift from multi-height observations will help reduce the noise in far-side maps, therefore enhancing the detectability of weaker active regions and helping modeling of the far-side heliosphere. ***Key observations to address this include multi-wavelength, multi-height measurements of the velocity and vector magnetic field that span from the photosphere to the base of the corona.***

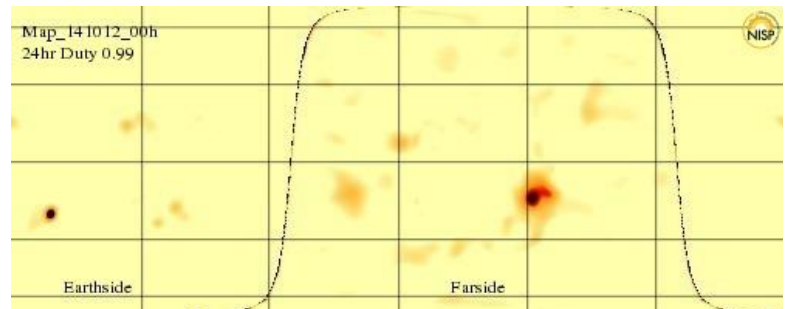

*Figure 2:* A map of the far-side of the Sun produced from helioseismic observations from GONG on Oct 12, 2014. The large prominent feature on the far-side is Active Region AR12192, the largest region in 25 years.

## Summary

The behavior of the Sun's magnetic field as it expands into the heliosphere is the driver of space-weather phenomena that impact all bodies in our Solar System; exoplanets orbiting magnetically active stars will experience similar phenomena. Within our planetary system, synoptic solar data hold the key to advancing our understanding of the structure and dynamics of the solar magnetic field and interior, and to provide input data for heliospheric models.

Key measurement capabilities for future ground-based solar synoptic facilities, involving the US and the broader international communities, include the following:

- Measure the boundary data as a function of height that propagate the magnetic connectivity from the solar surface into the heliosphere;
- Map the 3-D magnetic topology of solar erupting structures in the chromosphere and corona, increasing advanced warning of space-weather events from hours to days;
- Anticipate processes in the solar interior and the far side that impact heliospheric conditions; and
- Provide context for high-resolution observations of the Sun as well as for *in situ* single-point measurements throughout the heliosphere.